\documentclass[a4paper, mathleft]{an}
\Pagespan{1}{}
\Yearpublication{2012}
\Yearsubmission{2012}
\Month{1}
\Volume{333}
\Issue{10}
\DOI{10.1002/asna.2012xxxxx}

\usepackage{graphicx}
\usepackage{times}
\usepackage[usenames, dvipsnames]{color}
\usepackage{natbib}
\usepackage{amssymb}

\bibpunct{(}{)}{;}{a}{}{,}

\newcommand\tsp{\mbox{$\;\!$}}

\usepackage{hyperref}
\hypersetup{
    final=true,
    pageanchor=true,
    colorlinks=true,
    breaklinks=true,
    linkcolor=blue,
    citecolor=blue,
    urlcolor=blue,
    pdfpagemode=UseNone,
    pdftitle={A retrospective of the GREGOR solar telescope in scientific
        literature},
    pdfauthor={C. Denker et al.},
    pdfsubject={Solar Physics},
    pdfkeywords={telescopes -
        instrumentation: high angular resolution -
        instrumentation: adaptive optics -
        instrumentation: spectrographs -
        instrumentation: interferometers -
        instrumentation: polarimeters}}

\begin{document}

%
%

\title{A retrospective of the GREGOR solar telescope in scientific literature}
\author{C.~Denker\inst{1}\fnmsep\thanks{Corresponding author:
    \email{cdenker@aip.de}\newline},
    O.~von der L{\"u}he\inst{2},
    A.~Feller\inst{3},
    K.~Arlt\inst{1},
    H.~Balthasar\inst{1},
    S.-M.~Bauer\inst{1},
    N.~Bello Gonz\'alez\inst{2},
    T.~Berkefeld\inst{2},
    P.~Caligari\inst{2},
    M.~Collados\inst{4},
    A.~Fischer\inst{2},
    T.~Granzer\inst{2},
    T.~Hahn\inst{2},
    C.~Halbgewachs\inst{2},
    F.~Heidecke\inst{2},
    A.~Hofmann\inst{1},
    T.~Kentischer\inst{2},
    M.~Klva{\v n}a\inst{5},
    F.~Kneer\inst{6},
    A.~Lagg\inst{3},
    H.~Nicklas\inst{6},
    E.~Popow\inst{1},
    K.G.~Puschmann\inst{1},
    J.~Rendtel\inst{1},
    D.~Schmidt\inst{2},
    W.~Schmidt\inst{2},
    M.~Sobotka\inst{5},
    S.K.~Solanki\inst{3},
    D.~Soltau\inst{2},
    J.~Staude\inst{1},
    K.G.~Strassmeier\inst{1},
    R.~Volkmer\inst{2},
    T.~Waldmann\inst{2},
    E.~Wiehr\inst{6},
    A.D.~Wittmann\inst{6}, \and
    M.~Woche\inst{1}}
\titlerunning{The GREGOR solar telescope in scientific literature}
\authorrunning{C.~Denker et al.}
\institute{
    Leibniz-Institut f{\"u}r Astrophysik Potsdam,
    An der Sternwarte 16,
    14482 Potsdam,
    Germany
\and
    Kiepenheuer-Institut f{\"u}r Sonnenphysik,
    Sch{\"o}neckstra{\ss}e 6,
    79104 Freiburg,
    Germany
\and
    Max-Planck-Institut f{\"u}r Sonnensystemforschung,
    Max-Planck-Stra{\ss}e 2,
    37191 Katlenburg-Lindau, Germany
\and
    Instituto de Astrof\'{\i}sica de Canarias,
    C/ V\'{\i}a L\'actea s/n,
    38205 La Laguna,
    Tenerife,
    Spain
\and
    Astronomical Institute,
    Academy of Sciences of the Czech Republic,
    Fri{\v c}ova 298,
    25165 Ond{\v r}ejov, Czech Republic
\and
    Institut f\"ur Astrophysik,
    Georg-August-Universit{\"a}t G{\"o}ttingen,
    Friedrich-Hund-Platz 1,
    37077 G\"ottingen,
    Germany}
\received{18 Aug 2012}
\accepted{later}
\publonline{later}

%
%

\abstract{In this review, we look back upon the literature, which had the GREGOR
solar telescope project as its subject including science cases, telescope
subsystems, and post-focus instruments. The articles date back to the year 2000,
when the initial concepts for a new solar telescope on Tenerife were first
presented at scientific meetings. This comprehensive bibliography contains
literature until the year 2012, i.e., the final stages of commissioning and
science verification. Taking stock of the various publications in peer-reviewed
journals and conference proceedings also provides the ``historical'' context for
the reference articles in this special issue of \textit{Astronomische
Nachrichten/Astronomical Notes}.}
\maketitle

%
%

\section{GREGOR solar telescope project}

\begin{figure}[t]
\centerline{\includegraphics[width=\columnwidth]{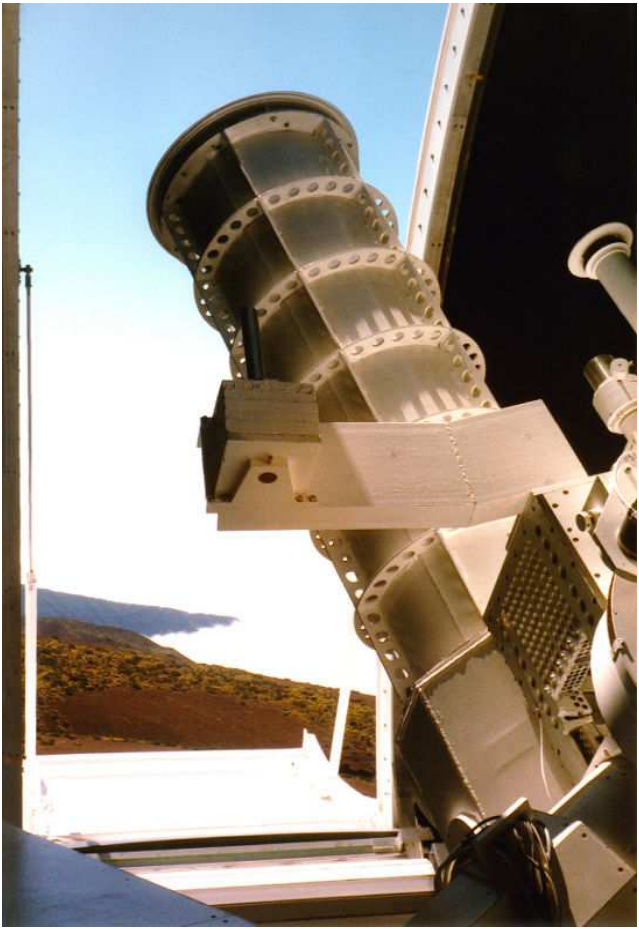}}
\caption{The Gregory-Coud\'e Telescope (GCT) at Observatorio del Teide, Iza\~na,
    Tenerife just after construction and commissioning in 1986
    (\textit{by courtesy of Axel Wittmann}). The venerable GCT has
    now been replaced by the GREGOR solar telescope.}
\label{FIG01}
\end{figure}

The GREGOR solar telescope \citep[see e.g.,][for early design
concepts]{vonderLuehe2000, vonderLuehe2001a, vonderLuehe2001b} supplanted the
Gregory-Coud{\'e} Telescope (GCT, see Fig.~\ref{FIG01}) at Observatorio del
Teide, Tenerife, Spain in 2002. In a brief history of the GCT, \cite{Wiehr2003}
emphasized the outstanding straylight properties of Gregory-type telescopes and
pointed to the GCT's excellent polarimetric properties, which are related to its
German-type coud\'e mounting and various optical means to compensate
instrumental polarization. The GREGOR project inherited the optical design that
had been developed for a scaled-down, open-telescope version of the Large
Earth-based Solar Telescope \citep[LEST,][]{Andersen2002}, while the LEST
foundation had to be resolved due to insufficient funding. The science
objectives as laid out in the original GREGOR proposal and recounted by
\citet{Schmidt2012a} are: (1) the interaction between convection and the
magnetic field in the photosphere, (2) solar magnetism and its role in driving
solar variability, (3) the enigmatic heating mechanisms of the chromosphere, and
(4) the search for solar twins during nighttime.

A series of review articles and status reports accompanied the construction and
commissioning phases of the GREGOR telescope \citep{Balthasar2007, Schmidt2012a,
Volkmer2003b, Volkmer2004, Volkmer2005, Volkmer2006, Volkmer2007, Volkmer2010a,
Volkmer2010b}. Substantial efforts were devoted to the design and construction
of the main optics from advanced Cesic\textsuperscript{\textregistered} ceramics
\citep[e.g.,][]{Volkmer2003b, Volkmer2006}. Problems in manufacturing such a
mirror ultimately led to a significant delay of the GREGOR project, which
finally reverted to a light-weighted Zerodur\textsuperscript{\textregistered}
mirror \citep{Volkmer2010a, Volkmer2010b}. The progress of the project can be
tracked in the reviews as critical milestones were reached: remodeling of the
GREGOR building, installation of the foldable tent dome, telescope control
systems \citep{Volkmer2004}; completion and installation of the telescope
structure with subsequent pointing, tracking, and thermal tests
\citep{Volkmer2005}; preparation of the optical laboratory for the adaptive
optics (AO) system and post-focus instruments, start of integration of the main
optics \citep{Volkmer2006}; installation of the interim 1-meter SolarLite mirror
and cooling system tests, integration of AO system and post-focus instruments,
start of commissioning \citep{Volkmer2010a, Volkmer2010b}. 

%
%

\section{Foldable tent dome}

The Dutch Open Telescope (DOT) already included many ingredients of the
``open principle'' \citep{Hammerschlag2009}, which were ultimately integrated
into the GREGOR telescope, i.e., wind-flushing of the main optical path, a
cooled field/heat stop in the primary focus, removal of all heat sources from
the observing deck, passive means to keep the telescope structure in thermal
equilibrium with its surroundings, a sturdy structure to prevent telescope shake
by winds, and absence of any large objects near the telescope.

The GREGOR foldable-tent dome consists of double membranes stretched between a
movable steel-bow structure \citep{Bettonvil2007, Bettonvil2008}. The
double-clothed tent improves thermal insulation and prevents condensation, while
the tension force stabilizes the structure when the two shells of dome are
closed. The dome has been in operation since 2006 and already survived major
storms with strong winds of up to 70~m~s$^{-1}$. The two halves of the dome can
be opened and closed at wind speeds of up to 30~m~s$^{-1}$. The smooth Teflon
coating of the polyvinylidene fluoride (PVDF) tent cloth minimizes snow and ice
loads.

The DOT and GREGOR domes are prototypes \citep{Hammerschlag2010,
Hammerschlag2012} envisioned for even larger constructions such as the European
Solar Telescope (EST). Therefore, both domes were equipped with a suite of
sensors, e.g., wind and pressure sensors as well as optical triangulation
sensors \citep{Jaeger2008} to measure deformations of the tent dome and its
performance in harsh/severe weather conditions. First results from the
three-dimensional dome displacement (3DD3) sensors were presented in
\citet{Sliepen2008}, which was followed up by a comprehensive analysis including
all sensors \citep{Sliepen2010}.

%
%

\section{Telescope structure}

\begin{figure}[t]
\centerline{\includegraphics[width=\columnwidth]{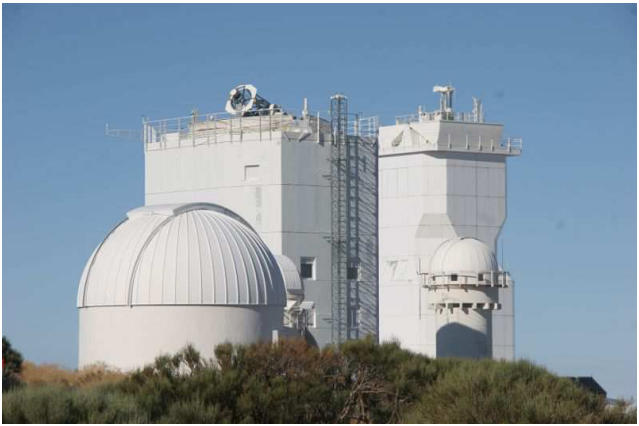}}
\caption{The 1.5-meter GREGOR solar telescope \mbox{(\textit{left}\tsp)} and the
    70-cm Vacuum Tower Telescope \mbox{(\textit{right}\tsp)} at Observatorio del
    Teide, Iza\~na, Tenerife (\textit{by courtesy of J\"urgen Rendtel}\tsp).}
\label{FIG02}
\end{figure}

The GREGOR telescope (see Fig.~\ref{FIG02}) with its open-truss steel structure
has to operate at wind speed of up to 20~m~s$^{-1}$, once the tent dome is
completely opened. In this environment, telescope seeing has to be avoided and
the influence of the wind blowing through the open telescope structure has to be
minimized. \citet{Emde2003} carried out finite element analyses and time-history
simulations of the structural response to static and dynamic wind loads. They
concluded that the absolute pointing accuracy is better than 1\arcsec\ (rms) for
all combinations of elevation and azimuth. Furthermore, the rms-wavefront errors
resulting from static and dynamic deformations of the telescope are less than
$\lambda / 10$.

%
%

\section{Thermal control}

An open-telescope design requires a strict control of the thermal environment.
The telescope structure should be within $-0.5$ to $+0.2$~K with the ambient
temperature, and the temperature differential of the primary mirror should be
less than 2~K. \citet{Emde2004} introduced the active and passive measures taken
to offset the deleterious effects of telescope and mirror seeing. These
precautions include sunshields to protect the telescope structure from direct
sunlight and surface coatings (TiO$_2$ paint and metallic foils). The thermal
control concept for telescope structure and primary mirror (air cooling from the
backside) were validated with finite element analyses \citep{Emde2004} and
direct as well as contactless measurements \citep{Volkmer2008}. The most recent
cooling system tests \citep{Volkmer2010a, Volkmer2010b} indicate only negligible
temperature differentials between ambient air and primary mirror.

%
%

\section{Telescope optics}
 
The light-weighted primary mirror made from
Zerodur\textsuperscript{\textregistered} ceramics has an aperture with a
diameter $D_1 = 1500$~mm. It is a paraboloid with a focal length of $f_1 =
2500$~mm \citep[e.g.,][]{vonderLuehe2001a}. The double Gregory configuration
with two elliptical Cesic\textsuperscript{\textregistered} mirrors ($D_2 =
400$~mm, $f_2 = 520$~mm, $D_3 = 300$~mm, and $f_3 = 1400$~mm) facilitates
placing a field stop in the primary focus and a polarization calibration unit in
the secondary focus. \citet{Volkmer2003b} illustrated the tight geometry of the
light beam in the vicinity of the heat stop (primary focus) and the polarization
calibration unit (secondary focus). The alignment of the secondary mirror with
respect to the primary mirror is critical for the optical quality of the
telescope. Therefore, the secondary mirror is mounted on a hexapod, where it can
be positioned with high precision in the micrometer and 10-$\mu$rad range.
Alignment constraints are less critical for the tertiary mirror, which is
mounted  behind the primary mirror and illuminated through a central hole with a
diameter of 380~mm. Accordingly, the tertiary is used for focusing the
telescope. The effective focal length of the telescope is $f_\mathrm{eff} =
58.950$~mm resulting in a focal ratio $f / D \approx 39$ and an image scale of
3.5\arcsec~mm$^{-1}$ in the science focus. A field-of-view (FOV) with a maximal
diameter of 300\arcsec\ can be observed. Currently, the heat stop restricts the
FOV to a circular area with a diameter of 180\arcsec. A comparison of the VTT
and GREGOR thoeretical modulation transfer functions was presented in
\citet{Soltau2003}.


\subsection{Ceramic mirrors}

The GREGOR telecope was the first major solar telescope condering
Cesic\textsuperscript{\textregistered} ceramics for its main optics. The initial
optical design envisioned that the primary, secondary, and tertiary mirrors were
all made from this silicon-carbide composite material. This choice was driven by
the desire to have a light-weight, stiff optics that can be effectively cooled
to avoid mirror seeing \citep[see][]{Volkmer2003a}. In the end, only the
secondary and tertiary mirrors used this advanced material. The manufacturing
process of the Cesic\textsuperscript{\textregistered} ceramic mirrors is
described in \citet{Kroedel2006}, who recount all manufacturing steps from
creating, light-weighting, milling, and fusing the greenbody segments -- over
Si-infiltration of the mirror blanks and coating of the mirror surfaces with a
Si-SiC slurry -- to finally, lapping, polishing, and testing of the mirrors.
Once it became clear that the Cesic\textsuperscript{\textregistered} technology
was not sufficiently mature to create a large-aperture primary mirror, a more
conventional approach was chosen using Zerodur\textsuperscript{\textregistered}
as the mirror substrate. The impacts on mirror design and support cell are laid
out in \citet{Suess2010}, whereas fabrication issues were presented by
\citet{Westerhoff2010}, who also discuss aggressively light-weighted designs for
mirrors with up to 4-m diameter.


\subsection{GREGOR polarimetric unit}

The polarimetric projects at the GREGOR solar telescope were summarized in
\citet{Hofmann2007, Hofmann2008}. The general formalism to calibrate the
instrumental polarization of the GREGOR telescope was set forth in
\citet{Balthasar2011}, which accounts for all mirrors of the alt-azimuthally
mounted telescope, the mirrors and vacuum windows of the coud\'e train, and an
optional image derotator. The goal is to reach a polarimetric sensitivity at the
$10^{-4}$ level in terms of the continuum intensity. The GREGOR Polarimetric
Unit \citep[GPU,][]{Hofmann2003} was developed to this end, serving all
post-focus instruments in the visible and near-infrared. The GPU is located in
the shadow of the Nasmyth mirror close to the secondary focus of the telescope,
where it calibrates all instrumental polarization occurring between the tertiary
mirror and the polarimeters of the post-focus instruments. The instrumental
polarization introduced by the on-axis primary and secondary mirrors is at or
below the $10^{-4}$ level and can be neglected. The GPU is also used to evaluate
the performance of the polarimeters used in the post-focus instruments. The GPU
consists of two motorized wheels, which contain positions for a pinhole, a
target, free apertures, and precision rotation mounts for a linear polarizer
(Marple-Hess prism) and polymethylmetcrylat (PMMA) waveplates. The GPU was
initially tested at the solar observatory ``Einsteinturm'' in Potsdam. The
procedures to derive the extinction ratios of the linear polarizer, the
wavelength dependency of the  retarders, and their efficiency to generate
circular polarized light were meticulously described in \citet{Hofmann2009}.


\subsection{Adaptive optics}

The scientific potential of modern solar telescope is substantially enhanced by
AO systems. The first design specifications for a ``first-light'' AO system were
laid out by \citet{vonderLuehe2002} aiming at a diffraction-limited performance
of the GREGOR solar telescope for the quarter of observing time with the best
seeing conditions. The final design evolved significantly over the years
\citep{Berkefeld2004, Berkefeld2005, Berkefeld2006, Berkefeld2012} and inherited
many features of the Kiepenheuer Adaptive Optics System (KAOS), which has been
reliably operated at the Vacuum Tower Telescope (VTT, see Fig.~\ref{FIG02}) for
many years. This decade-long activity to develop solar AO systems also spun off
to projects such as EST and the balloon-borne SUNRISE telescope
\citep{Berkefeld2010}. 

The GREGOR AO system uses a correlating Shack-Hartmann wavefront sensor. A
high-cadence camera and high-performance computer provide a 2.5-kHz control loop
frequency to achieve a 0-dB bandwidth of correction at about 130~Hz. The
correction of the image motion is off\-loaded to a separate tip-tilt mirror so
that the deformable mirror actuators can effectively correct the non-linear
wavefront aberrations. The optical design also includes an option for
multi-conjugate adaptive optics \citep[MCAO, e.g.,][]{Berkefeld2010,
Schmidt2009, Schmidt2010, Schmidt2012b}. The MCAO system has been extensively
tested at the Kiepenheuer-Institut and the VTT, which led to substantial
improvements in the optical design, namely the location of the on- and off-axis
wavefront sensor. Thus, intensity fluctuations of a few percent
(``flying-shadows'') caused by a warped pupil image could simply be removed by a
slightly undersized pupil stop. The controlled FOV of the MCAO system has a
diameter of about 60\arcsec, which is about ten times larger than what can be
achieved with conventional AO.

The primary mirror's small radius of curvature necessitates a high-degree of
control over the secondary that is mounted on a hexapod. The DC components of
the wavefront sensor signals encompass the wavefront aberrations of all optical
elements in front of the wavefront sensor. \citet{Soltau2006} described the use
of the AO system as commissioning tool for telescope alignment by solving the
inversion problem of extracting the alignment parameters from the response
matrix of the actuators.

%
%

\section{Post-focus instruments}

The initial plans for the GREGOR post-focus instrumentation
\citep[e.g.,][]{Kneer2001b} included proposals for visible-light and infrared
spectrographs, slit-jaw devices, polarimeters, broad-band imagers, and an
imaging spectropolarimeter. Only the visible-light Czerny-Turner spectrograph
and an adaption of the Polarimetric Littrow Spectrometer
\citep[POLIS,][]{Schmidt2003} were not included in the suite of ``first-light''
instruments. Dichroic pentaprisms are used to split the light between the
visible and infrared wavelengths. This photon-efficient set-up enables
simultaneous multi-wavelength observations with several instruments. Similarly,
a high-performance notch filter using thin-film optical coatings on a pentaprism
\citep{Schallenberg2010} sends only light within a band of 10--15~nm centered at
500~nm to the wavefront sensor of the AO system.


\subsection{Grating Infrared Spectrograph (GRIS)}

The Grating Infrared Spectrograph \citep[GRIS,][]{Collados2008} is a
Czerny-Turner spectrograph designed for high-spectral resolution (${\cal R}
\approx$ 300,000--450,000) observations in the wavelength range from
1.0--1.8~$\mu$m, e.g., the Si\,\textsc{i} $\lambda 1.079$~$\mu$m, He\,\textsc{i}
$\lambda 1.083$~$\mu$m, and Fe\,\textsc{i} $\lambda 1.565$~$\mu$m ($g = 3$)
spectral lines. The spectrograph can be combined with the Tenerife Infrared
Polarimeter (TIP-2). The slit-length is 70\arcsec\ and the image scale amounts
to 0.135\arcsec\ pixel$^{-1}$. Scanning a FOV of $70\arcsec \times 70\arcsec$
will take less than 20~min while recording full-Stokes spectra with
signal-to-noise ratio of 1000:1. GRIS reaches the diffraction-limit of the
telescope at $\lambda 1.565$~$\mu$m. Multi-wavelength observations are enabled
by slit-jaw cameras and in combination with the GREGOR Fabry-P\'erot
Interferometer (GFPI).


\subsection{GREGOR Fabry-P\'erot Interferometer (GFPI)}

Conceptually, Fabry-P{\'e}rot interferometers can be placed in a telecentric or
collimated mounting. \citet{Kneer2001a} discussed advantages and disadvantages
of this choice and presented the first general ideas for an imaging spectrometer
at the GREGOR solar telescope. The final optical layout of the GFPI
\citep{Denker2010b, Puschmann2012a, Puschmann2012b} can be found in
\citet{Puschmann2007}. The new instrument is based on the ``G\"ottingen''
Fabry-P\'erot Interferometer, which received upgrades of the etalons, cameras
and software in 2005 \citep{Puschmann2006} and was outfitted with a polarimeter
based on ferro-electric liquid crystals and a modified Savart plate in 2007
\citep{BelloGonzalez2008, Balthasar2009}. The GFPI was transferred from the VTT
to the GREGOR solar telescope in 2009. The spectral resolution of the GFPI is
about ${\cal R} \approx$ 250,000 in the wavelength range from 530--860~nm. Two
spectral lines can be observed sequentially while covering a FOV of $52\arcsec
\times 40\arcsec$. A second Fabry-P\'erot interferometer for the blue spectral
region was proposed by \citet{Denker2010a}, who addressed the opportunities but
also the challenges for imaging spectropolarimetry brought forth by new
large-format, high-cadence camera systems.


\subsection{Broad-Band Imager (BBI)}

A recently installed Broad-Band Imager (BBI) records large-format, high-cadence
images, which are suitable for image restoration. \citet{Schmidt2012a} presented
its optical layout including a Foucault channel to measure the optical quality
and performance of the telescope.


\subsection{Gregor@night}

Nighttime operation was foreseen since the inception of the GREGOR telescope
\citep{Volkmer2003a}. The initial plans included a fiber-fed echelle double
spectrograph with a spectral resolution of ${\cal R} \approx$ 100,000 covering
two wavelength regions at 360--490~nm and 510--870~nm \citep{Strassmeier2007}.
The unrestricted availability of the GREGOR telescope at night makes it an ideal
choice for surveys of solar-type stars, the search for solar twins, and
monitoring of stellar activity cycles (e.g., in the Ca\,\textsc{ii}\,H\,\&\,K
lines). Current plans call for an adaptation of the SOFIN spectrograph of the
Nordic Optical Telescope (NOT).

%
%

\section{Conclusions}

The design, construction, and commissioning phases of the GREGOR project were
described in about 60 articles in refereed journals and conference proceedings.
This article summarizes this body of work and serves as a reference to this
special issue of \textit{Astro\-no\-mische Nach\-rich\-ten/Astro\-nomical
Notes}, which sums up the state of the GREGOR project at the end of the science
verification phase in 2012.

%
%

\acknowledgements The 1.5-meter GREGOR solar telescope was build by a German
consortium under the leadership of the Kiepenheuer-Institut f{\"u}r Sonnenphysik
in Freiburg with the Leibniz-Institut f{\"u}r Astrophysik Potsdam, the Institut
f{\"u}r Astrophysik G{\"o}ttingen, and the Max-Planck-Institut f{\"u}r
Sonnensystemforschung in Katlenburg-Lindau as partners, and with contributions
by the Instituto de Astrof{\'{\i}}sica de Canarias and the Astronomical
Institute of the Academy of Sciences of the Czech Republic. CD was supported by
grant DE~787/3-1 of the Deutsche Forschungsgemeinschaft (DFG).

%
%


\begin{thebibliography}{59}
\bibitem[\protect\citeauthoryear{Andersen, Engvold \&
Owner-Petersen}{2002}]{Andersen2002}
    Andersen, T., Engvold, O., Owner-Petersen, M.: 2002, \textit{LEST
    Technical Report No.~64}, Institute for Theoretical Astrophysics, University
    of Oslo
\bibitem[\protect\citeauthoryear{Balthasar et al.}{2007}]{Balthasar2007}
    Balthasar, H., von der L{\"u}he, O., Kneer, F., et al.: 2007, in: P.\
    Heinzel, I.\ Dorotovi{\v c}, R.J.\ Rutten (eds.), \textit{The Physics of
    Chromospheric Plasmas}, ASPC 368, p.~605
\bibitem[\protect\citeauthoryear{Balthasar et al.}{2009}]{Balthasar2009}
    Balthasar, H., Bello Gonz{\'a}lez, N., Collados, M., et al.: 2009, in: K.G.\
    Strassmeier, A.G.\ Kosovichev, J.E.\ Beckmann (eds.), \textit{Cosmic
    Magnetic Fields: From Planets, to Stars and Galaxies}, IAU Symp.\ 259,
    p.~665
\bibitem[\protect\citeauthoryear{Balthasar et al.}{2011}]{Balthasar2011}
    Balthasar, H., Bello Gonz{\'a}lez, N., Collados, M., et al.: 2011, in: J.R.\
    Kuhn et al.\ (eds.), \textit{Solar Polarization Workshop 6}, ASPC 437,
    p.~351
\bibitem[\protect\citeauthoryear{Bello Gonz{\'a}lez \&
    Kneer}{2008}]{BelloGonzalez2008}
    Bello Gonz{\'a}lez, N., Kneer, F.: 2008, A\&A 480, 265
\bibitem[\protect\citeauthoryear{Berkefeld, Soltau \& von der
    L{\"u}he}{2004}]{Berkefeld2004}
    Berkefeld, T., Soltau, D., von der L{\"u}he, O.: 2004, in: D.\ Bonaccini
    Calia, B.L.\ Ellerbroek, R.\ Ragazzoni (eds.), \textit{Advancements in
    Adaptive Optics}, SPIE 5490, p.~260
\bibitem[\protect\citeauthoryear{Berkefeld, Soltau \& von der
    L{\"u}he}{2005}]{Berkefeld2005}
    Berkefeld, T., Soltau, D., von der L{\"u}he, O.: 2005, in: R.K.\ Tyson, M.\
    Lloyd-Hart (eds.), \textit{Astronomical Adaptive Optics Systems and
    Applications II}, SPIE 5903, p.~219
\bibitem[\protect\citeauthoryear{Berkefeld, Soltau \& von der
    L{\"u}he}{2006}]{Berkefeld2006}
    Berkefeld, T., Soltau, D., von der L{\"u}he, O.: 2006, in: B.L.\ Ellerbroek,
    D.\ Bonaccini Calia (eds.), \textit{Advances in Adaptive Optics II}, SPIE
    6272, p.~05
\bibitem[\protect\citeauthoryear{Berkefeld et al.}{2010}]{Berkefeld2010}
    Berkefeld, T., Soltau, D., Schmidt, D., von der L{\"u}he, O.: 2010, Appl.\
    Opt.\ 49, G155
\bibitem[\protect\citeauthoryear{Berkefeld et al.}{2012}]{Berkefeld2012}
    Berkefeld, T., Schmidt, D., Soltau, D., et al.: 2012, in: Ellerbroek, B.L.,
    Marchetti, E., V{\'e}ran, J.P. (eds.), \textit{Adaptive Optics Systems III},
    SPIE 8447, p.~51
\bibitem[\protect\citeauthoryear{Bettonvil et al.}{2007}]{Bettonvil2007}
    Bettonvil, F.C.M., Hammerschlag, R.H., J{\"a}gers, A.P.L., Sliepen, G.:
    2007, in: \textit{Structural Architecture. Toward the Future Looking to the
    Past}, Venice, p.~204
\bibitem[\protect\citeauthoryear{Bettonvil et al.}{2008}]{Bettonvil2008}
    Bettonvil, F.C.M., Hammerschlag, R.H., J{\"a}gers, A.P.L., Sliepen, G.:
    2008, in: E.\ Atad-Ettedgui, D.\ Lemke (eds.), \textit{Advanced Optical and
    Mechanical Technologies in Telescopes and Instrumentation}, SPIE 7018,
    p.~1N
\bibitem[\protect\citeauthoryear{Collados et al.}{2008}]{Collados2008}
    Collados, M., Calcines, A., D{\'{\i}}az, J.J., et al: 2008, in: I.S.\
    McLean, M.M.\ Casali (eds.), \textit{Ground-Based and Airborne
    Instrumentation for Astronomy II}, SPIE 7014, p.~5Z
\bibitem[\protect\citeauthoryear{Denker}{2010}]{Denker2010a}
    Denker, C.: 2010, AN 331, 648
\bibitem[\protect\citeauthoryear{Denker et al.}{2010}]{Denker2010b}
    Denker, C., Balthasar, H., Hofmann, A., et al.: 2010, in: I.S.\ McLean,
    S.K.\ Ramsay, H.\ Takami (eds.), \textit{Ground-Based and Airborne
    Instrumentation for Astronomy III}, SPIE 7735, p.~6M
\bibitem[\protect\citeauthoryear{Emde et al.}{2003}]{Emde2003}
    Emde, P., S{\"u}{\ss}, M., Eisentr{\"a}ger, P., et al.: 2003, in: W.A.\
    Goodman (ed.), \textit{Optical Materials and Structures Technologies}, SPIE
    5179, p.~282
\bibitem[\protect\citeauthoryear{Emde et al.}{2004}]{Emde2004}
    Emde, P., K{\"u}hn, J., Weis, U., Bornkessel, T.: 2004, in: J.\ Antebi, D.\
    Lemke (eds.), \textit{Astronomical Structures and Mechanisms
    Technology}, SPIE 5495, p.~238
\bibitem[\protect\citeauthoryear{Hammerschlag et al.}{2009}]{Hammerschlag2009}
    Hammerschlag, R.H., Bettonvil, F.C.M., J{\"a}gers, A.P.L., Sliepen, G.:
    2009, Earth, Moon, \& Planets 104, 83
\bibitem[\protect\citeauthoryear{Hammerschlag et al.}{2010}]{Hammerschlag2010}
    Hammerschlag, R.H., Kommers, J.N.M., van Leverink, S.J., et al.: 2010, in:
    L.M.\ Stepp, R.\ Gilmozzi, H.J.\ Hall (eds.), \textit{Ground-Based and
    Airborne Telescopes III}, SPIE 7733, p.~0J
\bibitem[\protect\citeauthoryear{Hammerschlag et al.}{2012}]{Hammerschlag2012}
    Hammerschlag, R.H., Kommers, J.N., Visser, S., et al.: 2012, in: Navarro,
    R., Cunningham, C.R., Pietro, E. (eds.), \textit{Modern Technologies in
    Space-and Ground-Based Telescopes and Instrumentation II}, SPIE 8450, p.~07
\bibitem[\protect\citeauthoryear{Hofmann}{2007}]{Hofmann2007}
    Hofmann, A.: 2007, Sun \& Geosphere 2, 9
\bibitem[\protect\citeauthoryear{Hofmann}{2008}]{Hofmann2008}
    Hofmann, A.: 2008, Centr.\ Eur.\ Astrophys.\ Bull.\ 32, 17
\bibitem[\protect\citeauthoryear{Hofmann \& Rendtel}{2003}]{Hofmann2003}
    Hofmann, A., Rendtel, J.: 2003, in: S.\ Fineschi (ed.), \textit{Polarimetry
    in Astronomy}, SPIE 4843, p.~112
\bibitem[\protect\citeauthoryear{Hofmann, Rendtel \& Arlt}{2009}]{Hofmann2009}
    Hofmann, A., Rendtel, J., Arlt, K.: 2009, Centr.\ Eur.\ Astrophys.\ Bull.\
    33, 317
\bibitem[\protect\citeauthoryear{J{\"a}gers et al.}{2008}]{Jaeger2008}
    J{\"a}gers, A.P.L., Sliepen, G., Bettonvil, F.C.M., Hammerschlag, R.H.:
    2008, in: E.\ Atad-Ettedgui, D.\ Lemke (eds.), \textit{Advanced Optical and
    Mechanical Technologies in Telescopes and Instrumentation}, SPIE 7018,
    p.~1R
\bibitem[\protect\citeauthoryear{Kneer \& Hirzberger}{2001}]{Kneer2001a}
    Kneer, F., Hirzberger, H.: 2001, AN 322, 375
\bibitem[\protect\citeauthoryear{Kneer et al.}{2001}]{Kneer2001b}
    Kneer, F., Hofmann, A., von der L{\"u}he, O., et al.: 2001, AN 322, 361
\bibitem[\protect\citeauthoryear{Kr{\"o}del, Luichtel \&
    Volkmer}{2006}]{Kroedel2006}
    Kr{\"o}del, M.R., Luichtel, G., Volkmer, R.: 2006, in: E.\ Atad-Ettedgui,
    D.\ Lemke (eds.), \textit{Optomechanical Technologies for Astronomy}, SPIE
    6273, p.~0Q
\bibitem[\protect\citeauthoryear{Puschmann et al.}{2006}]{Puschmann2006}
    Puschmann, K.G., Kneer, F., Seelemann, T., Wittmann, A.D.: 2006, A\&A 451,
    1151
\bibitem[\protect\citeauthoryear{Puschmann et al.}{2007}]{Puschmann2007}
    Puschmann, K.G., Kneer, F., Nicklas, H., Wittmann, A.D.: 2007, in: F.\
    Kneer, K.G.\ Puschmann, A.D.\ Wittmann (eds.), \textit{Modern Solar
    Facilities -- Advanced Solar Science}, p.~45
\bibitem[\protect\citeauthoryear{Puschmann et al.}{2012a}]{Puschmann2012a}
    Puschmann, K.G., Balthasar, H., Bauer, S.-M., et al.: 2012a, in:
    \textit{Magnetic Fields from the Photosphere to the Corona}, Rimmele,
    T.R.\ (ed.), ASPC, ArXiv e-prints, 1111.5509
\bibitem[\protect\citeauthoryear{Puschmann}{2012b}]{Puschmann2012b}
    Puschmann, K.G., Balthasar, H., Popow, E.,et al.: 2012b, in: McLean, I.S.,
    Ramsay, S.K., Takami, H. (eds.), \textit{Ground-Based and Airborne
    Instrumentation for Astronomy IV}, SPIE 8446, p.~276
\bibitem[\protect\citeauthoryear{Schallenberg et al.}{2010}]{Schallenberg2010}
    Schallenberg, U., Ploss, B., Lappschies, M., Jakobs, S.: 2010, in: E.\
    Atad-Ettedgui, D.\ Lemke (eds.), \textit{Modern Technologies in Space- and
    Ground-Based Telescopes and Instrumentation}, SPIE 7739, p.~1X
\bibitem[\protect\citeauthoryear{Schmidt et al.}{2009}]{Schmidt2009}
    Schmidt, D., Berkefeld, T., Heidecke, F., et al: 2009, in: P.G.\ Warren et
    al.\ (eds.), \textit{Astronomical and Space Optical Systems}, SPIE 7439,
    p.~0X
\bibitem[\protect\citeauthoryear{Schmidt et al.}{2010}]{Schmidt2010}
    Schmidt, D., Berkefeld, T., Feger, B., Heidecke, F.: 2010, in: B.L.\
    Ellerbroek et al.\ (eds.), \textit{Adaptive Optics Systems II}, SPIE 7736,
    p.~07
\bibitem[\protect\citeauthoryear{Schmidt \& Berkefeld}{2012}]{Schmidt2012b}
    Schmidt, D., Berkefeld, T.: 2012, in: Ellerbroek, B.L., Marchetti, E.,
    V{\'e}ran, J.P. (eds.), \textit{Adaptive Optics Systems III}, SPIE 8447,
    p.~128
\bibitem[\protect\citeauthoryear{Schmidt et al.}{2003}]{Schmidt2003}
    Schmidt, W., Beck, C., Kentischer, T., et al.: 2003, AN 324, 300
\bibitem[\protect\citeauthoryear{Schmidt et al.}{2012}]{Schmidt2012a}
    Schmidt, W., von der L{\"u}he, O., Volkmer, R., et al.: 2012, in:
    \textit{Magnetic Fields from the Photosphere to the Corona}, Rimmele,
    T.R.\ (ed.), ASPC, ArXiv e-prints, 1202.4289
\bibitem[\protect\citeauthoryear{Sliepen et al.}{2008}]{Sliepen2008}
    Sliepen, G., J{\"a}gers, A.P.L., Bettonvil, F.C.M., Hammerschlag, R.H.:
    2008, in: E.\ Atad-Ettedgui, D.\ Lemke (eds.), \textit{Advanced Optical and
    Mechanical Technologies in Telescopes and Instrumentation}, SPIE 7018,
    p.~1C
\bibitem[\protect\citeauthoryear{Sliepen et al.}{2010}]{Sliepen2010}
    Sliepen, G., J{\"a}gers, A.P.L., Hammerschlag, R.H., Bettonvil, F.C.M.:
    2010, in: L.M.\ Stepp, R.\ Gilmozzi, H.J.\ Hall (eds.), \textit{Ground-Based
    and Airborne Telescopes III}, SPIE 7733, p.~32
\bibitem[\protect\citeauthoryear{Soltau, Berkefeld \&
    Volkmer}{2006}]{Soltau2006}
    Soltau, D., Berkefeld, T., Volkmer, R.: 2006, in: L.M.\ Stepp (ed.),
    \textit{Ground-Based and Airborne Telescopes}, SPIE 6267, p.~11
\bibitem[\protect\citeauthoryear{Soltau et al.}{2003}]{Soltau2003}
    Soltau, D., Berkefeld, T., von der L{\"u}he, O., et al.: 2003, AN 324, 292
\bibitem[\protect\citeauthoryear{Strassmeier et al.}{2007}]{Strassmeier2007}
    Strassmeier, K.G., Woche, M., Granzer, T., et al.: 2007, in: F.\ Kneer,
    K.G.\ Puschmann, A.D.\ Wittmann (eds.), \textit{Modern Solar Facilities --
    Advanced Solar Science}, p.~51
\bibitem[\protect\citeauthoryear{S{\"u}{\ss}, Volkmer, \&
    Eisentr{\"a}ger}{2010}]{Suess2010}
    S{\"u}{\ss}, M., Volkmer, R., Eisentr{\"a}ger, P.: 2010, in: E.\
    Atad-Ettedgui, D.\ Lemke (eds.), \textit{Modern Technologies in Space- and
    Ground-Based Telescopes and Instrumentation}, SPIE 7739, p.~1I
\bibitem[\protect\citeauthoryear{Volkmer}{2008}]{Volkmer2008}
    Volkmer, R.: 2008, in: L.M.\ Stepp, R.\ Gilmozzi (eds.),
    \textit{Ground-Based and Airborne Telescopes II}, SPIE 7012, p.~0K
\bibitem[\protect\citeauthoryear{Volkmer et al.}{2003a}]{Volkmer2003a}
    Volkmer, R., von der L{\"u}he, O., Soltau, D., et al.: 2003a, in: W.A.\
    Goodman (ed.), \textit{Optical Materials and Structures Technologies}, SPIE
    5179, p.~270
\bibitem[\protect\citeauthoryear{Volkmer et al.}{2003b}]{Volkmer2003b}
    Volkmer, R., von der L{\"u}he, O., Kneer, F.: 2003b, in: S.K.\ Keil, S.V.\
    Avakyan (eds.), \textit{Innovative Telescopes and Instrumentation for Solar
    Astrophysics}, SPIE 4853, p.~360
\bibitem[\protect\citeauthoryear{Volkmer et al.}{2004}]{Volkmer2004}
    Volkmer, R., von der L{\"u}he, O., Kneer, F., et al.: 2004, in: J.M.\
    Oschmann (ed.), \textit{Ground-Based Telescopes}, SPIE 5489, p.~693
\bibitem[\protect\citeauthoryear{Volkmer et al.}{2005}]{Volkmer2005}
    Volkmer, R., von der L{\"u}he, O., Kneer, F., et al.: 2005, in: S.\
    Fineschi, R.A.\ Viereck  (eds.),  \textit{Solar Physics and Space Weather
    Instrumentation}, SPIE 5901, p.~75
\bibitem[\protect\citeauthoryear{Volkmer et al.}{2006}]{Volkmer2006}
   Volkmer, R., von der L{\"u}he, O., Kneer, F., et al.: 2006, in: L.M.\ Stepp
   (ed.), \textit{Ground-Based and Airborne Telescopes}, SPIE 6267, p.~0W
\bibitem[\protect\citeauthoryear{Volkmer et al.}{2007}]{Volkmer2007}
    Volkmer, R., von der L{\"u}he, O., Kneer, F., et al.: 2007, in: F.\ Kneer,
    K.G.\ Puschmann, A.D.\ Wittmann (eds.), \textit{Modern Solar Facilities --
    Advanced Solar Science}, p.~39
\bibitem[\protect\citeauthoryear{Volkmer et al.}{2010a}]{Volkmer2010a}
    Volkmer, R., von der L{\"u}he, Denker, C., et al.: 2010a, AN 331, 624
\bibitem[\protect\citeauthoryear{Volkmer et al.}{2010b}]{Volkmer2010b}
    Volkmer, R., von der L{\"u}he, Denker, C., et al.: 2010b, in: L.M.\ Stepp,
    R.\ Gilmozzi, H.J.\ Hall (eds.), \textit{Ground-Based and Airborne
    Telescopes III}, SPIE 7733, p.~0K
\bibitem[\protect\citeauthoryear{von der L{\"u}he, Berkefeld \&
    Soltau}{2002}]{vonderLuehe2002}
    von der L{\"u}he, O., Berkefeld, T., Soltau, D.: 2002, in: A.\ Kohnle, J.D.\
    Gonglewski, T.J.\ Schmugge (eds.), \textit{Optics in Atmospheric Propagation
    and Adaptive Systems IV}, SPIE 4538, p.~197
\bibitem[\protect\citeauthoryear{von der L{\"u}he et
    al.}{2000}]{vonderLuehe2000}
    von der L{\"u}he, O., Schmidt, W., Soltau, D., et al.: 2000, in: A.\ Wilson
    (ed.), \textit{The Solar Cycle and Terrestrial Climate, Solar and Space
    Weather}, ESA SP 463, p.~629
\bibitem[\protect\citeauthoryear{von der L{\"u}he et
    al.}{2001a}]{vonderLuehe2001a}
    von der L{\"u}he, O., Schmidt, W., Soltau, D., et al.: 2001a, AN 322, 353
\bibitem[\protect\citeauthoryear{von der L{\"u}he et
    al.}{2001b}]{vonderLuehe2001b}
    von der L{\"u}he, O., Schmidt, W., Soltau, D., et al.: 2001b, in: B.\
    Battrick et al.\ (eds.), \textit{Solar Encounter}, ESA SP 493, p.~417
\bibitem[\protect\citeauthoryear{Westerhoff et al.}{2010}]{Westerhoff2010}
    Westerhoff, T., Sch{\"a}fer, M., Thomas, A., et al.: 2010, in: E.\
    Atad-Ettedgui, D.\ Lemke (eds.), \textit{Modern Technologies in Space- and
    Ground-Based Telescopes and Instrumentation}, SPIE 7739, p.~0M
\bibitem[\protect\citeauthoryear{Wiehr}{2003}]{Wiehr2003}
    Wiehr, E.: 2003, AN 324, 285
\end{thebibliography}
\end{document}